\documentclass[aps,prl,twocolumn,showpacs,superscriptaddress,groupedaddress]{revtex4}

\usepackage{graphicx}
\usepackage{dcolumn}
\usepackage{bm}
\usepackage{amssymb}

\usepackage[usenames]{color}

\begin{document}

\newcommand{\met}       {\mbox{$\not\!\!E_T$}}
\newcommand{\deta}      {\mbox{$\eta^{\rm det}$}}
\newcommand{\meta}      {\mbox{$\left|\eta\right|$}}
\newcommand{\mdeta}     {\mbox{$\left|\eta^{\rm det}\right|$}}
\newcommand{\rar}       {\rightarrow}
\newcommand{\rargap}    {\mbox{ $\rightarrow$ }}
\newcommand{\ttbar}     {\mbox{$t\bar{t}$}}
\newcommand{\bbbar}     {\mbox{$b\bar{b}$}}
\newcommand{\ccbar}     {\mbox{$c\bar{c}$}}
\newcommand{\qqbar}     {\mbox{$q\bar{q}$}}
\newcommand{\ppbar}     {\mbox{$p\bar{p}$}}
\newcommand{\comphep}   {\sc{c}\rm{omp}\sc{hep}}
\newcommand{\herwig}    {\sc{herwig}}
\newcommand{\pythia}    {\sc{pythia}}
\newcommand{\alpgen}    {\sc{alpgen}}
\newcommand{\singletop} {\rm{SingleTop}}
\newcommand{\reco}      {\sc{reco}}
\newcommand{\Mchiggs}   {\mbox{$M({\rm jet1,jet2},W)$}}
\newcommand{\coss}	{\mbox{\rm{cos}$\theta^{\star}$}}
\newcommand{\ljets} {$\ell+$jets}

\lefthyphenmin=6
\righthyphenmin=6

\hspace{5.2in}\mbox{Fermilab-Pub-10/203-E}

\title{Search for flavor changing neutral currents via quark-gluon couplings in single top quark 
production using 2.3~fb$^{-1}$ of $\bm{p\bar{p}}$~collisions
}\
\affiliation{Universidad de Buenos Aires, Buenos Aires, Argentina}
\affiliation{LAFEX, Centro Brasileiro de Pesquisas F{\'\i}sicas, Rio de Janeiro, Brazil}
\affiliation{Universidade do Estado do Rio de Janeiro, Rio de Janeiro, Brazil}
\affiliation{Universidade Federal do ABC, Santo Andr\'e, Brazil}
\affiliation{Instituto de F\'{\i}sica Te\'orica, Universidade Estadual Paulista, S\~ao Paulo, Brazil}
\affiliation{Simon Fraser University, Vancouver, British Columbia, and York University, Toronto, Ontario, Canada}
\affiliation{University of Science and Technology of China, Hefei, People's Republic of China}
\affiliation{Universidad de los Andes, Bogot\'{a}, Colombia}
\affiliation{Charles University, Faculty of Mathematics and Physics, Center for Particle Physics, Prague, Czech Republic}
\affiliation{Czech Technical University in Prague, Prague, Czech Republic}
\affiliation{Center for Particle Physics, Institute of Physics, Academy of Sciences of the Czech Republic, Prague, Czech Republic}
\affiliation{Universidad San Francisco de Quito, Quito, Ecuador}
\affiliation{LPC, Universit\'e Blaise Pascal, CNRS/IN2P3, Clermont, France}
\affiliation{LPSC, Universit\'e Joseph Fourier Grenoble 1, CNRS/IN2P3, Institut National Polytechnique de Grenoble, Grenoble, France}
\affiliation{CPPM, Aix-Marseille Universit\'e, CNRS/IN2P3, Marseille, France}
\affiliation{LAL, Universit\'e Paris-Sud, CNRS/IN2P3, Orsay, France}
\affiliation{LPNHE, Universit\'es Paris VI and VII, CNRS/IN2P3, Paris, France}
\affiliation{CEA, Irfu, SPP, Saclay, France}
\affiliation{IPHC, Universit\'e de Strasbourg, CNRS/IN2P3, Strasbourg, France}
\affiliation{IPNL, Universit\'e Lyon 1, CNRS/IN2P3, Villeurbanne, France and Universit\'e de Lyon, Lyon, France}
\affiliation{III. Physikalisches Institut A, RWTH Aachen University, Aachen, Germany}
\affiliation{Physikalisches Institut, Universit{\"a}t Freiburg, Freiburg, Germany}
\affiliation{II. Physikalisches Institut, Georg-August-Universit{\"a}t G\"ottingen, G\"ottingen, Germany}
\affiliation{Institut f{\"u}r Physik, Universit{\"a}t Mainz, Mainz, Germany}
\affiliation{Ludwig-Maximilians-Universit{\"a}t M{\"u}nchen, M{\"u}nchen, Germany}
\affiliation{Fachbereich Physik, Bergische  Universit{\"a}t Wuppertal, Wuppertal, Germany}
\affiliation{Panjab University, Chandigarh, India}
\affiliation{Delhi University, Delhi, India}
\affiliation{Tata Institute of Fundamental Research, Mumbai, India}
\affiliation{University College Dublin, Dublin, Ireland}
\affiliation{Korea Detector Laboratory, Korea University, Seoul, Korea}
\affiliation{CINVESTAV, Mexico City, Mexico}
\affiliation{FOM-Institute NIKHEF and University of Amsterdam/NIKHEF, Amsterdam, The Netherlands}
\affiliation{Radboud University Nijmegen/NIKHEF, Nijmegen, The Netherlands}
\affiliation{Joint Institute for Nuclear Research, Dubna, Russia}
\affiliation{Institute for Theoretical and Experimental Physics, Moscow, Russia}
\affiliation{Moscow State University, Moscow, Russia}
\affiliation{Institute for High Energy Physics, Protvino, Russia}
\affiliation{Petersburg Nuclear Physics Institute, St. Petersburg, Russia}
\affiliation{Stockholm University, Stockholm and Uppsala University, Uppsala, Sweden }
\affiliation{Lancaster University, Lancaster LA1 4YB, United Kingdom}
\affiliation{Imperial College London, London SW7 2AZ, United Kingdom}
\affiliation{The University of Manchester, Manchester M13 9PL, United Kingdom}
\affiliation{University of Arizona, Tucson, Arizona 85721, USA}
\affiliation{University of California Riverside, Riverside, California 92521, USA}
\affiliation{Florida State University, Tallahassee, Florida 32306, USA}
\affiliation{Fermi National Accelerator Laboratory, Batavia, Illinois 60510, USA}
\affiliation{University of Illinois at Chicago, Chicago, Illinois 60607, USA}
\affiliation{Northern Illinois University, DeKalb, Illinois 60115, USA}
\affiliation{Northwestern University, Evanston, Illinois 60208, USA}
\affiliation{Indiana University, Bloomington, Indiana 47405, USA}
\affiliation{Purdue University Calumet, Hammond, Indiana 46323, USA}
\affiliation{University of Notre Dame, Notre Dame, Indiana 46556, USA}
\affiliation{Iowa State University, Ames, Iowa 50011, USA}
\affiliation{University of Kansas, Lawrence, Kansas 66045, USA}
\affiliation{Kansas State University, Manhattan, Kansas 66506, USA}
\affiliation{Louisiana Tech University, Ruston, Louisiana 71272, USA}
\affiliation{University of Maryland, College Park, Maryland 20742, USA}
\affiliation{Boston University, Boston, Massachusetts 02215, USA}
\affiliation{Northeastern University, Boston, Massachusetts 02115, USA}
\affiliation{University of Michigan, Ann Arbor, Michigan 48109, USA}
\affiliation{Michigan State University, East Lansing, Michigan 48824, USA}
\affiliation{University of Mississippi, University, Mississippi 38677, USA}
\affiliation{University of Nebraska, Lincoln, Nebraska 68588, USA}
\affiliation{Rutgers University, Piscataway, New Jersey 08855, USA}
\affiliation{Princeton University, Princeton, New Jersey 08544, USA}
\affiliation{State University of New York, Buffalo, New York 14260, USA}
\affiliation{Columbia University, New York, New York 10027, USA}
\affiliation{University of Rochester, Rochester, New York 14627, USA}
\affiliation{State University of New York, Stony Brook, New York 11794, USA}
\affiliation{Brookhaven National Laboratory, Upton, New York 11973, USA}
\affiliation{Langston University, Langston, Oklahoma 73050, USA}
\affiliation{University of Oklahoma, Norman, Oklahoma 73019, USA}
\affiliation{Oklahoma State University, Stillwater, Oklahoma 74078, USA}
\affiliation{Brown University, Providence, Rhode Island 02912, USA}
\affiliation{University of Texas, Arlington, Texas 76019, USA}
\affiliation{Southern Methodist University, Dallas, Texas 75275, USA}
\affiliation{Rice University, Houston, Texas 77005, USA}
\affiliation{University of Virginia, Charlottesville, Virginia 22901, USA}
\affiliation{University of Washington, Seattle, Washington 98195, USA}
\author{V.M.~Abazov} \affiliation{Joint Institute for Nuclear Research, Dubna, Russia}
\author{B.~Abbott} \affiliation{University of Oklahoma, Norman, Oklahoma 73019, USA}
\author{M.~Abolins} \affiliation{Michigan State University, East Lansing, Michigan 48824, USA}
\author{B.S.~Acharya} \affiliation{Tata Institute of Fundamental Research, Mumbai, India}
\author{M.~Adams} \affiliation{University of Illinois at Chicago, Chicago, Illinois 60607, USA}
\author{T.~Adams} \affiliation{Florida State University, Tallahassee, Florida 32306, USA}
\author{G.D.~Alexeev} \affiliation{Joint Institute for Nuclear Research, Dubna, Russia}
\author{G.~Alkhazov} \affiliation{Petersburg Nuclear Physics Institute, St. Petersburg, Russia}
\author{A.~Alton$^{a}$} \affiliation{University of Michigan, Ann Arbor, Michigan 48109, USA}
\author{G.~Alverson} \affiliation{Northeastern University, Boston, Massachusetts 02115, USA}
% ...
%
%
\collaboration{The D0 Collaboration\footnote{with visitors from
$^{a}$Augustana College, Sioux Falls, SD, USA,
$^{b}$The University of Liverpool, Liverpool, UK,
$^{c}$SLAC, Menlo Park, CA, USA,
$^{d}$ICREA/IFAE, Barcelona, Spain,
$^{e}$Centro de Investigacion en Computacion - IPN, Mexico City, Mexico,
$^{f}$ECFM, Universidad Autonoma de Sinaloa, Culiac\'an, Mexico,
and 
$^{g}$Universit{\"a}t Bern, Bern, Switzerland.%
}} \noaffiliation
\vskip 0.25cm

\date{June 17, 2010}

\begin{abstract}
We present a search for flavor changing neutral currents via quark-gluon couplings in a
sample of single top quark final states corresponding to 2.3~fb$^{-1}$ of integrated luminosity
collected with the D0~detector at the Fermilab Tevatron Collider.
We select events containing a single top quark candidates with an additional jet, and 
obtain separation between signal and background using Bayesian neural networks. We find consistency
between background expectation and observed data, and set limits on
flavor changing neutral current gluon couplings of the top quark to up~quarks 
($tgu$) and charm quarks ($tgc$).
The cross section limits at the $95\%$ C.L. are
 $\sigma_{tgu}< 0.20$~pb 
and 
$\sigma_{tgc}<0.27$~pb.
These correspond to limits on the top quark
decay branching fractions of  
$\mathcal{B}(t \rightarrow gu)<2.0\times10^{-4}$ and $\mathcal{B}(t \rightarrow gc)<3.9\times10^{-3}$.
\end{abstract}

\pacs{14.65.Ha, 12.15.Ji, 13.85.Qk}

\maketitle

The observation of electroweak production of single top quarks was reported in 
2009~\cite{Abazov:2009ii,Aaltonen:2009jj}. Those measurements, as well as the 
evidence for $t$-channel single top quark production~\cite{Abazov:2009pa}, focus on
standard model (SM) production of single top quarks. The single top
quark final state is sensitive to several models of physics beyond the SM~\cite{Tait:2000sh}, 
in particular 
those in which flavor changing neutral current (FCNC) couplings between a gluon, a top quark, 
and up or charm quarks may be large. Examples include models with multiple Higgs 
doublets such as supersymmetry~\cite{ss1,ss2,ss3}, those with new dynamical interactions of the 
top quark~\cite{tc1,tc2,Zhang:2007ub}, or models in which the top quark is a composite object~\cite{composite}
or has a soliton structure~\cite{soliton1,soliton2}. In principle, such interactions can also be produced 
through SM higher-order radiative corrections; however, their effects are too small to be 
observed~\cite{ss1}. Stringent limits exist for FCNC top quark couplings to photons and 
$Z$~bosons from studies of production and decay of top 
quarks~\cite{cdfrun1_fcnc,:2008aaa,lep_fcnc,hera_fcnc}. 
The first limits on gluon FCNC couplings to the top quark were obtained in a D0 analysis based on 
0.23~fb$^{-1}$ of integrated luminosity~\cite{Abazov:2007ev}. The CDF Collaboration searched for
FCNC production of single top quarks, without extra jets, through gluon-quark 
interactions~\cite{Yang:2006gs,Gao:2009rf}, using a dataset corresponding to 
2.2~fb$^{-1}$ of integrated luminosity~\cite{Aaltonen:2008qr}. The limits on the FCNC couplings
are $\kappa_{tgu}/\Lambda < 0.018\; {\rm TeV^{-1}}$ 
and 
$\kappa_{tgc}/\Lambda < 0.069\; {\rm TeV^{-1}}$, where $\Lambda$ is the scale of the 
new interactions which generate these couplings (of order 1~TeV).

\begin{figure}[!h!tbp]
\vspace{-0.07in}
\includegraphics[width=0.499\textwidth]{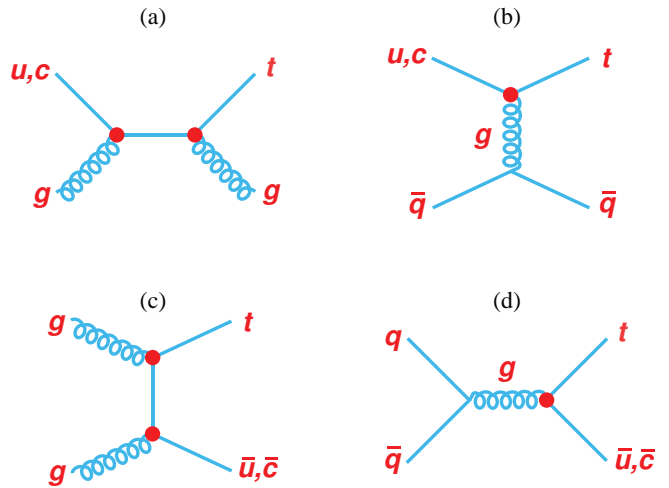}
\vspace{-0.1in}
\caption{Leading order Feynman diagrams for FCNC
gluon coupling between an up or a charm quark and a top quark.(a) and (d) show two $s$-channel 
diagrams for the $tg$ final state and the $tq$ final state and (b) and (c) are two
$t$-channel diagrams for the $tq$ final state. The circles indicate the effective FCNC
coupling, possible at either of the two vertices in (a) and (c), for which the amplitudes are properly summed.
\label{fig:feynman} }
\end{figure}

The FCNC coupling of a gluon to a top quark and a light quark results in either 
$s$-channel production and decay [Fig.~\ref{fig:feynman}~(a) and~(d)]
or in $t$-channel exchange [Fig.~\ref{fig:feynman}~(b) and~(c)] of a virtual particle. 
The largest contribution to the production cross section (83\% for $tgu$ and 66\% for $tgc$) 
is from the diagram in Fig.~\ref{fig:feynman}~(a).
The final state in each case contains a top quark and a light quark or gluon, a topology
similar to SM $t$-channel single top quark production. We do not consider the single top quark
final state without extra jets that was explored by the CDF Collaboration~\cite{Aaltonen:2008qr} 
due to its different final state topology and significantly smaller signal event yield for a given coupling.
~

The FCNC couplings can be parametrized in a model-independent way, using an 
effective Lagrangian~\cite{lagran1, lagran2} of the form: 
\begin{equation}
\label{eq:lagrangian}
{\cal L}_{\rm FCNC} = \frac{\kappa_{tgf}}{\Lambda} g_s \bar{f} \sigma^{\mu\nu} \frac{\lambda^a}{2} t G^a_{\mu\nu} ,
\end{equation}
where $f$ = $u$ or $c$, with $u$, $c$ and $t$ representing the quark fields; $\kappa_{tgf}$ defines 
the strength of the $tgu$ or $tgc$ couplings; $g_s$ and $\lambda^a$ are the strong coupling constant 
and color matrices; $\sigma^{\mu\nu}$ and $G^a_{\mu\nu}$ are the Dirac tensor and the gauge field 
tensor of the gluon. The FCNC single top quark production cross section depends therefore 
quadratically on the factor $\kappa_{tgf}/\Lambda$.
For a coupling of $\kappa_{tgf}/\Lambda=0.015$~TeV$^{-1}$, the next-to leading order (NLO) cross sections at 
a top quark mass of 170~GeV are
$\sigma^{\rm NLO}_{tgu}=0.29$~pb and $\sigma^{\rm NLO}_{tgc}=0.020$~pb~\cite{Liu}. 
The top quark decay branching 
fraction to a gluon and any quark also depends quadratically on the factor 
$\kappa_{tgf}/\Lambda$~\cite{Zhang:2008yn}, but this branching fraction is negligible for coupling factors 
considered in this analysis~\cite{lagran2}.

We search for FCNC production of single top quarks in association with a quark or gluon,
where the top quark decays to a $W$~boson and a $b$~quark, and the $W$~boson subsequently decays to a
lepton (electron or muon) and a neutrino. The main backgrounds
to this final state are from $W$+jets production, 
including $W+c$-quarks and $W+b$-quarks, with smaller contributions
from {\ttbar}, SM single top quarks ($tb+tqb$), as well as multijets, dibosons, and $Z$+jets production. 
We base the analysis 
on the dataset and event selection from the single top quark production observation 
Letter~\cite{Abazov:2009ii},
using 2.3~fb$^{-1}$ of integrated luminosity collected with the 
D0 detector~\cite{D0detector} at the Fermilab Tevatron Collider.

The detector has a central tracking system, consisting of a 
silicon microstrip tracker and a central fiber tracker, 
both located within a 1.9~T superconducting solenoidal 
magnet, with designs optimized for tracking and 
vertexing at pseudorapidities $|\eta|<3$ and $|\eta|<2.5$, respectively~\cite{D0SMTNIM,D0L0NIM,etaphi}.
A liquid-argon and uranium calorimeter has a 
central section covering pseudorapidities $|\eta|$ up to 
$\approx 1.1$, and two end calorimeters that extend coverage 
to $|\eta|\approx 4.2$, with all three housed in separate 
cryostats~\cite{run1det}. An outer muon system, at $|\eta|<2$, 
consists of a layer of tracking detectors and scintillation trigger 
counters in front of 1.8~T toroids, followed by two similar layers 
after the toroids~\cite{run2muon}. 

%
We select events containing a lepton, missing transverse energy (\met),
and two to four jets with transverse momentum $p_T>15$~GeV
and $|\eta|<3.4$ (allowing for jets from gluon radiation), 
with the leading (highest $p_T$) jet additionally satisfying $p_T>25$~GeV~\cite{jet-definition}. 
We require $20 < {\met} < 200$~GeV for events with two jets and 
$25 < {\met} < 200$~GeV for
events with three or four jets. Events must contain only one isolated
electron with $p_T>15$~GeV and $|\eta|<1.1$ ($p_T>20$~GeV for three-
or four-jet events), or one isolated muon with $p_T>15$~GeV and
$|\eta|<2.0$. The multijets background, where a jet is misidentified as an isolated lepton, is kept to
approximately~5\% of the total background by requiring the scalar sum of all transverse energies,
$H_T({\rm lepton},{\met},{\rm alljets})$, to be greater than 110 to 160~GeV, depending on
the lepton flavor and jet multiplicity, and by requiring that the {\met} is not colinear
with the axes of the lepton or the leading jet in the transverse plane. 
To enhance the fraction of top quark events, one of the
jets is required to be identified as originating from $b$~quark fragmentation through a neural network (NN) 
$b$-tagging algorithm~\cite{btagging-scanlon}. 
To partially reject background from $W+b\bar{b}$, {\ttbar}, and SM single top quark events,
each event is required to contain only one $b$-tagged jet (vetoing double-tagged events),
in contrast to SM single top quark analyses where double-tagged events are also considered.

We model the FCNC signals and SM single top quark background with the 
{\sc SingleTop} Monte Carlo (MC) generator~\cite{singletop-mcgen},
using CTEQ6M parton distribution functions~\cite{cteq1,cteq2}.
The {\alpgen} leading-order MC event 
generator~\cite{alpgen}, interfaced to {\pythia} for showering and hadronization~\cite{pythia},
is used to model {\ttbar}, $W$+jets, and $Z$+jets 
background, while {\pythia} is used to model diboson ($WW$, $WZ$ and $ZZ$) production.
We set the mass of the top quark to 170~GeV, and use the CTEQ6L1 parton distribution 
functions~\cite{cteq1,cteq2}. We use {\sc geant}~\cite{geant} to simulate
the response of the D0 detector to MC events. 
To model the effects of multiple interactions and detector noise, data from random $p\bar{p}$
crossings are overlaid on MC events.
The SM single top quark, {\ttbar}, diboson and $Z$+jets 
backgrounds are normalized to their predicted 
cross sections~\cite{singletop-xsec-kidonakis,ttbar-xsec,mcfm}. 
The $W$+jets background normalization and jet angular distributions are obtained from 
data samples without $b$-tagging requirements, and its flavor
composition is determined from data samples with different numbers of $b$-tagged jets.
We model the background from multijets production using data containing lepton candidates that fail 
one of the lepton identification requirements, but otherwise resemble the signal events.
In the muon channel, where a secondary muon in a jet is misidentified as an isolated muon,
this is accomplished by reversing the tight isolation criterion, 
whereas in the electron channel, where a jet is misidentified as an electron, we reverse the tight 
electron identification criteria~\cite{d0-prd-2008,Heinson:2010xh}.

We select a total of 3735 lepton+jets data events with only one $b$-tagged jet. 
The sample composition is given in Table~\ref{tab:eventyields}.
\begin{table}[!h!btp]
\centering
\vspace{-0.15in}
\caption{Event yields with uncertainty for each jet multiplicity for the electron and muon channels combined. 
The FCNC signals are each normalized to their observed cross section upper limits. The uncertainty
on the total background includes correlations amongst sources.}
\label{tab:eventyields}
\begin{tabular}
      {l@{\extracolsep{\fill}}r@{\extracolsep{0pt}$\:\pm\:$}l@{}%
        @{\extracolsep{\fill}}r@{\extracolsep{0pt}$\,\pm\,$}l@{}%
        @{\extracolsep{\fill}}r@{\extracolsep{0pt}$\,\pm\,$}l@{}}
\hline
\hline
Source & \multicolumn{2}{c}{2 jets} & \multicolumn{2}{c}{3 jets}
                                        & \multicolumn{2}{c}{4 jets}
\\
\hline
FCNC signal \\
~~$tgu$                   &   34  &   4  &   16 &  3 &   5 &  1 \\
~~$tgc$                   &   54  &   7  &   23 &  4 &   7 &  2 \\ 
Background\\
~~$W$+jets                &  1660 & 146  &  560 & 54 & 154 & 15 \\
~~$Z$+jets and dibosons   &   204 &  34  &   72 & 14 &  22 &  6 \\
~~SM single top           &   112 &  15  &   46 &  7 &  14 &  3 \\
~~\ttbar                  &   152 &  24  &  277 & 42 & 278 & 41 \\
~~Multijets               &   184 &  47  &   66 & 15 &  27 &  5 \\ 
Total background          &  2312 & 170  & \hspace{9pt}1021 & 84 & \hspace{9pt}495 & 53 \\
Data                      &  \multicolumn{2}{c}{2277}  
                          &  \multicolumn{2}{c}{ 958} 
                          &  \multicolumn{2}{c}{ 500}  \\
\hline
\hline
\end{tabular}
\vspace{-0.1in}
\end{table}

We further improve the sensitivity to FCNC through an application of 
Bayesian neural networks (BNN)~\cite{bayesianNNs,bnn-tanasijczuk,d0-prd-2008},
with settings identical to those detailed in Ref.~\cite{Abazov:2009ii}. A BNN is an
average over many individual neural networks~\cite{neuralnetworks-bishop} 
(100 networks are used in this analysis), where the parameters
for each network are sampled from the Bayesian posterior density distribution of the entire
network parameter space.

\begin{figure}[!h!tbp]
\centering
\includegraphics[width=0.238\textwidth]{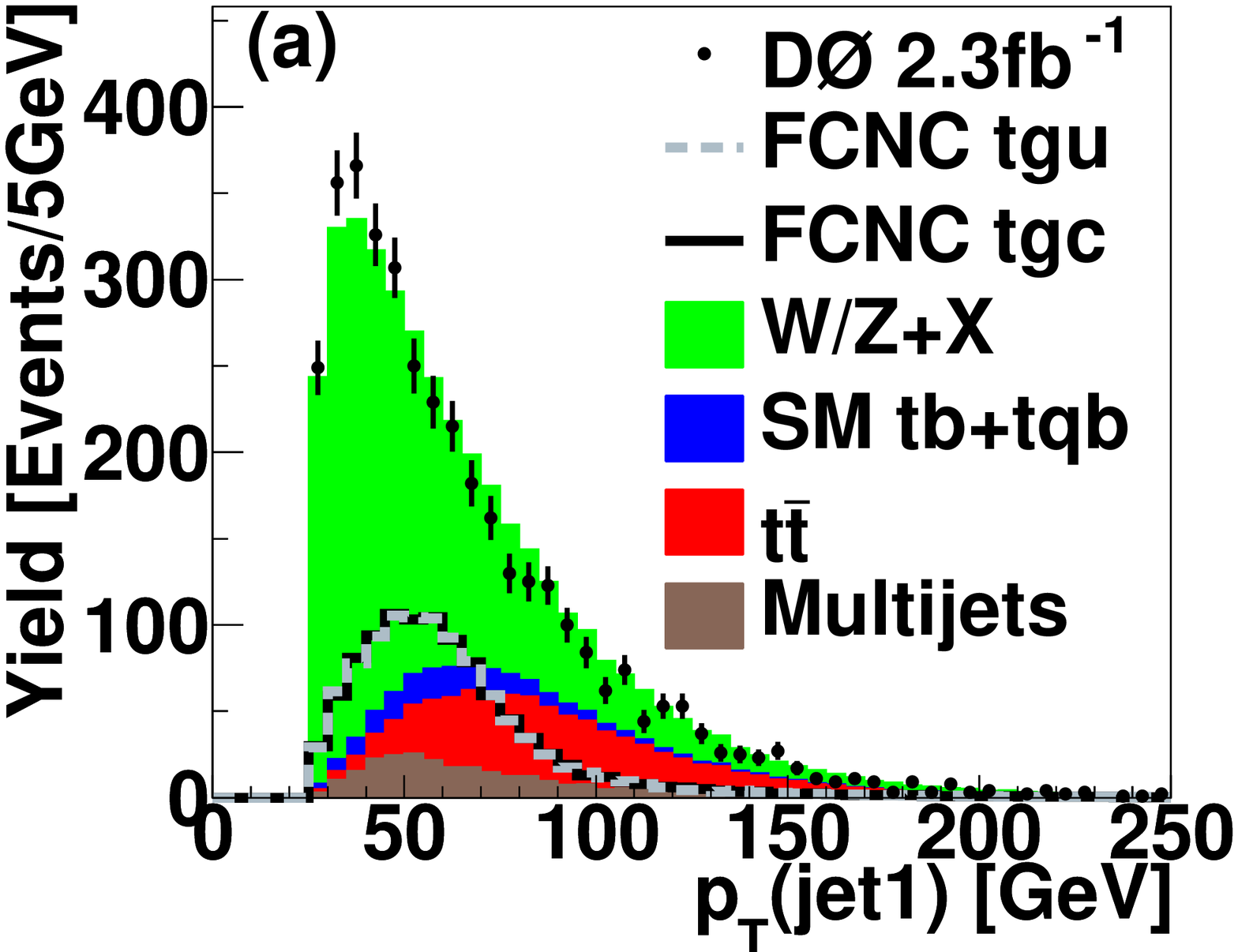}
\includegraphics[width=0.238\textwidth]{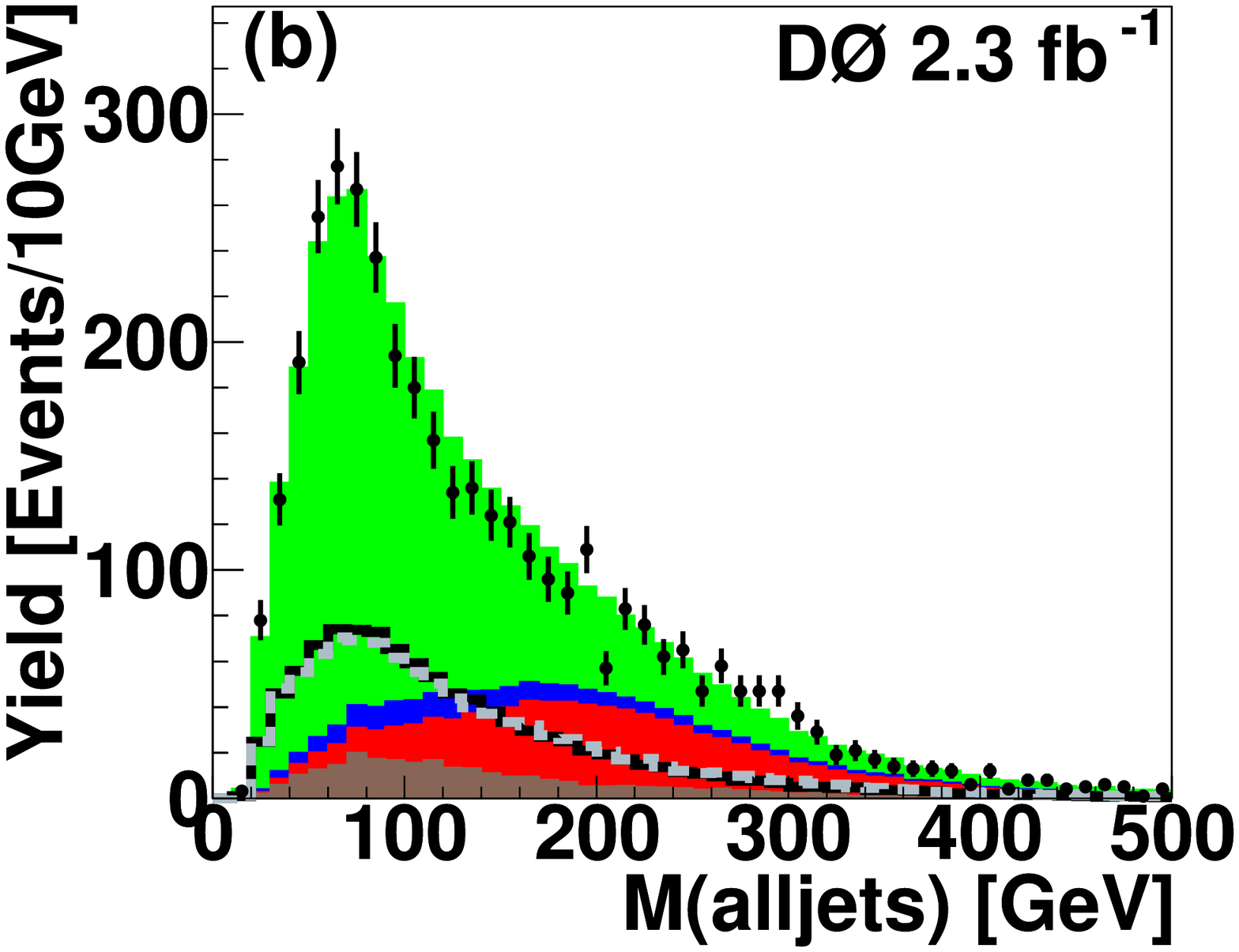}
\includegraphics[width=0.238\textwidth]{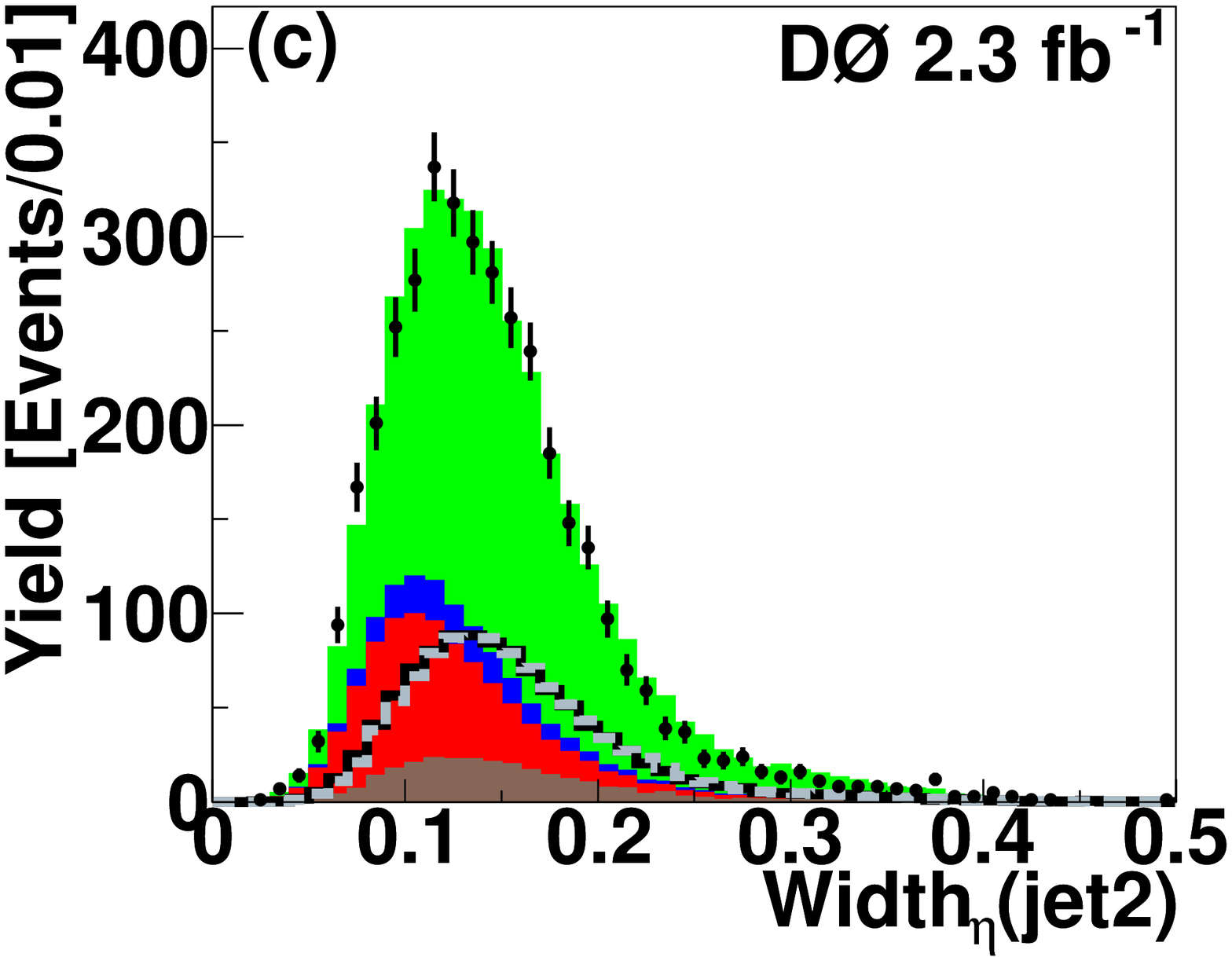}
\includegraphics[width=0.238\textwidth]{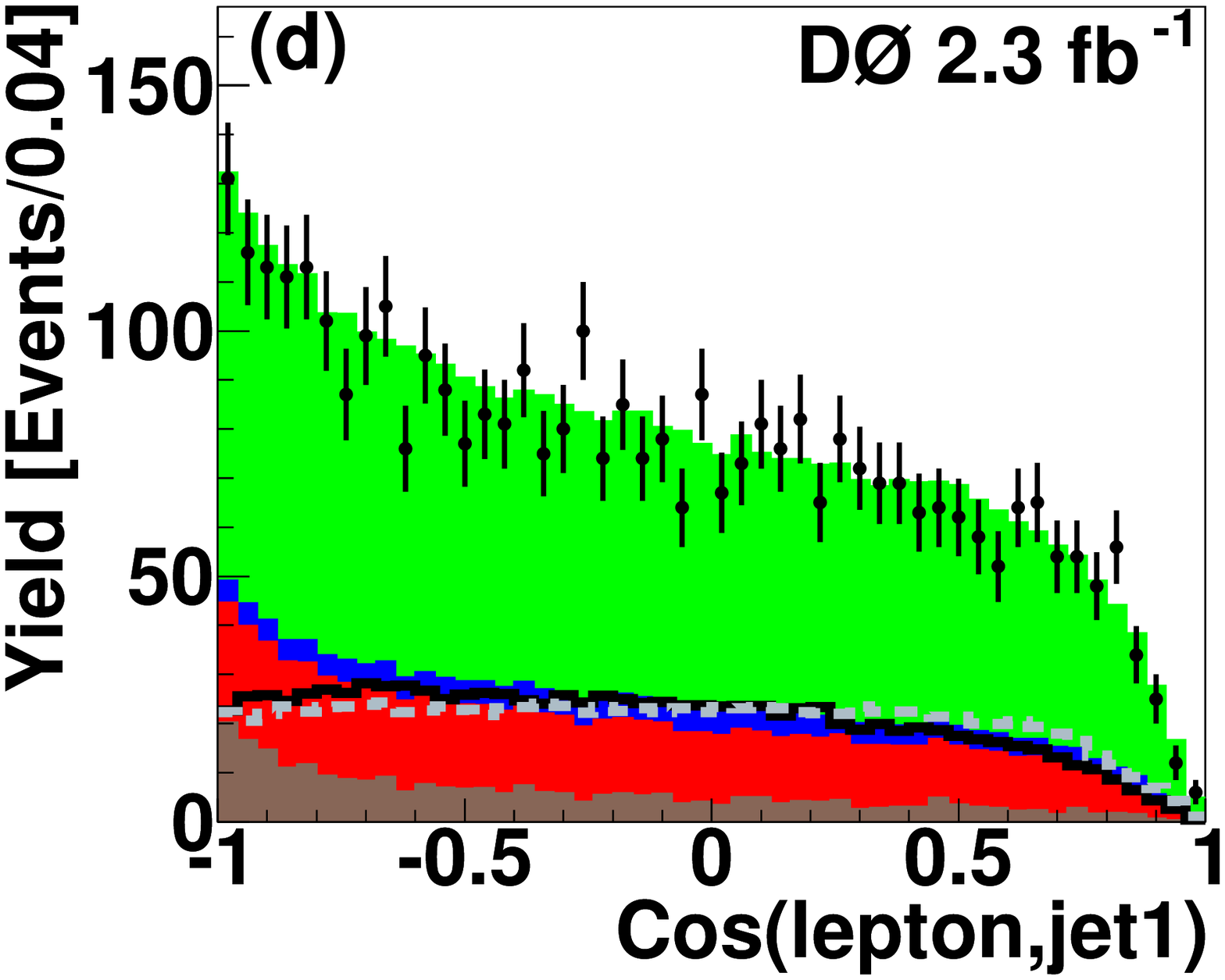}
\includegraphics[width=0.238\textwidth]{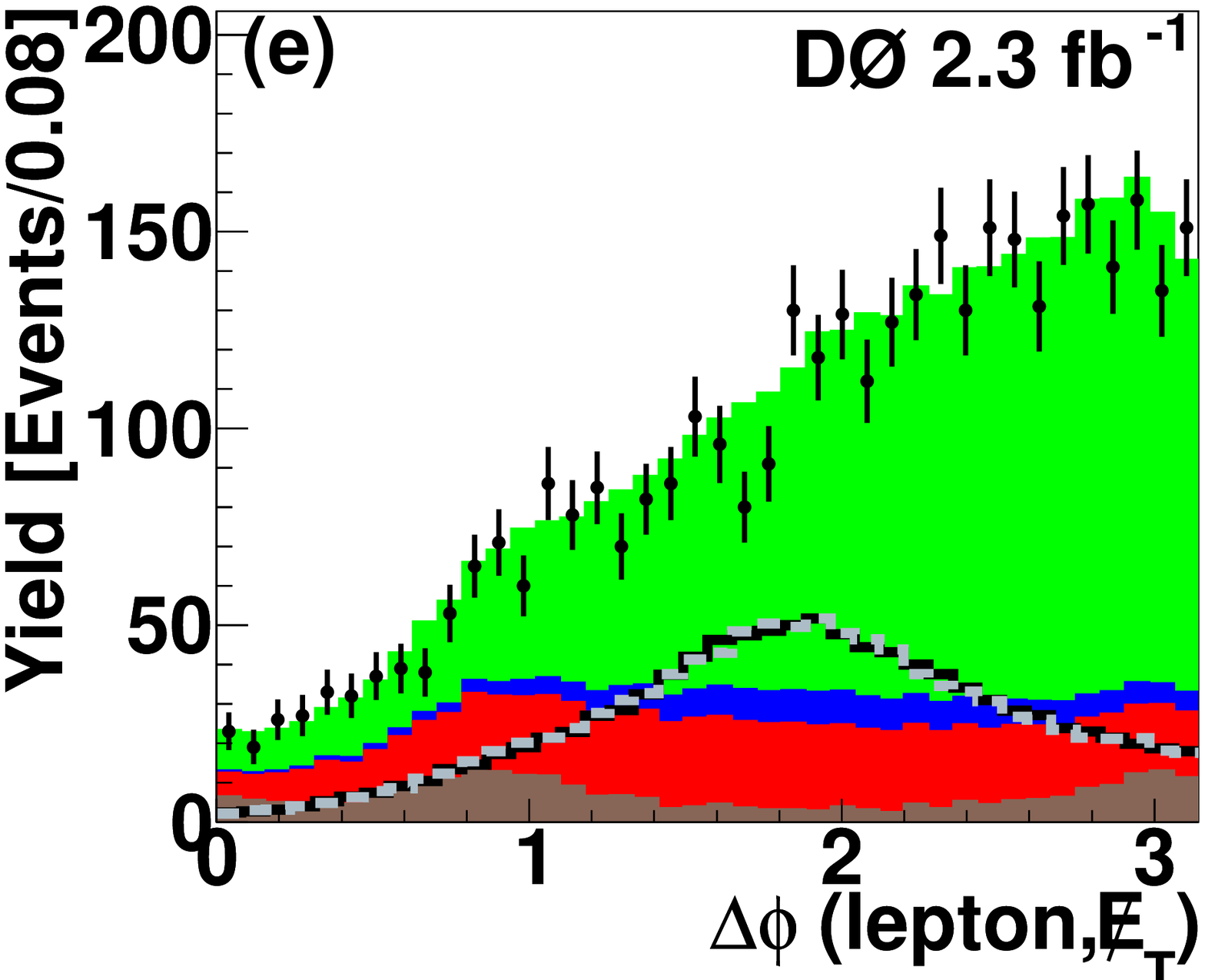}
\includegraphics[width=0.238\textwidth]{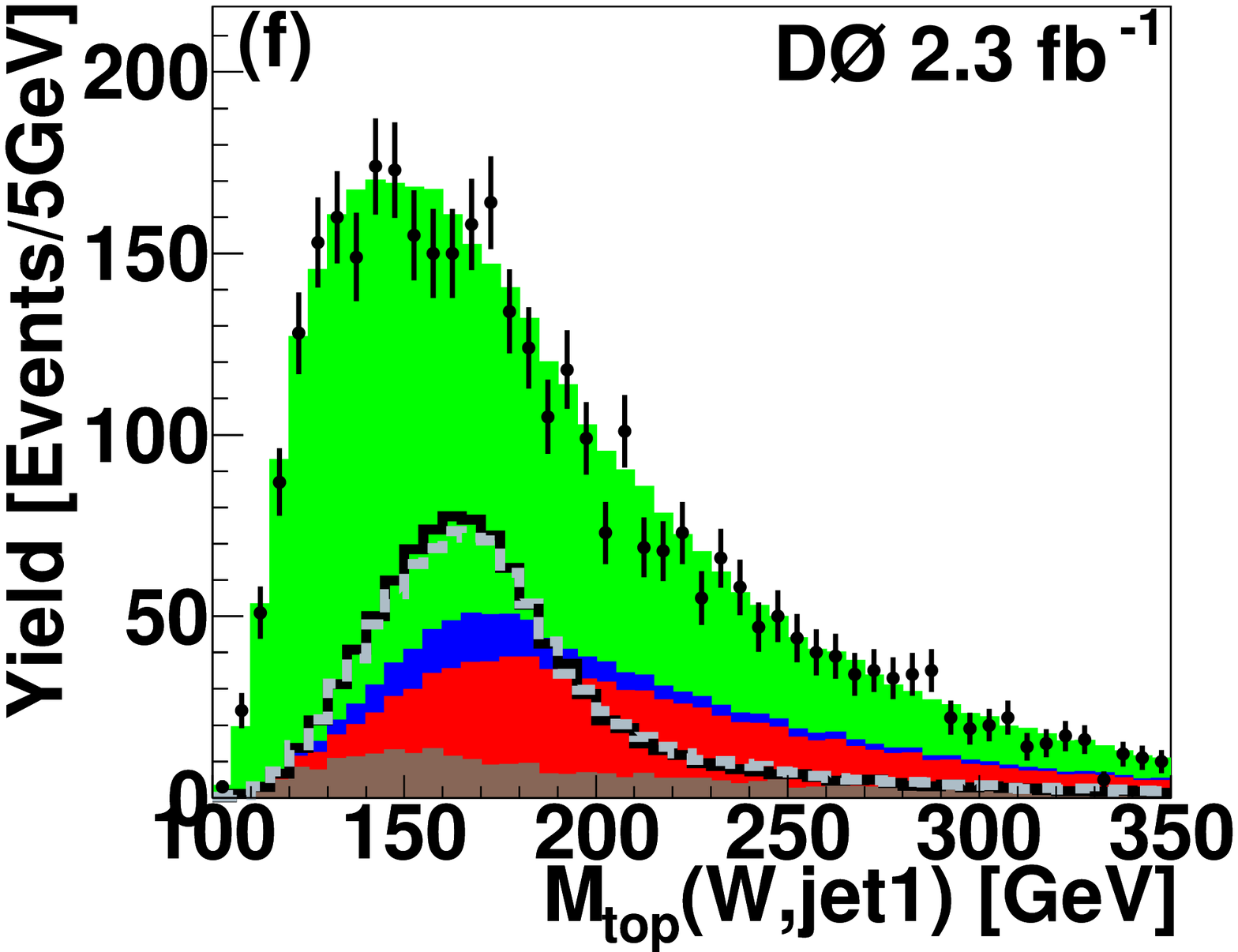}
\vspace{-0.1in}
\caption{Comparison of the background model to data for several discriminating variables summed over all
analysis channels: (a) $p_T$ of the leading jet, (b) invariant mass of the system of all jets, (c)
width in pseudorapidity of the second leading jet, (d) cosine of the angle between the leading jet and the 
lepton, (e) $\phi$ separation between the lepton and $\met$, and (f) top quark mass reconstructed from the
reconstructed $W$~boson and the leading jet. The FCNC signals are normalized to cross sections of 5~pb to 
visualize them clearly, and $W/Z+X$ includes $W$+jets and smaller backgrounds from $Z$+jets and dibosons.
\label{fig:vars} }
\vspace{-0.05in}
\end{figure}

We use 54 discriminating
variables, a subset of those used in each channel of the single top quark observation analysis~\cite{Abazov:2009ii} 
plus those from the previous FCNC analysis~\cite{Abazov:2007ev}. 
The set of variables comprises individual object and event kinematics, top quark reconstruction, 
jet width, and angular correlations. 
Figure~\ref{fig:vars} compares the observed data to the background model for six illustrative discriminating
variables. Object kinematics, such as the leading jet $p_T$, and event
kinematics, such as the invariant mass of the all-jets system, help
separate the FCNC signals from the $W$+jets background. Jet
reconstruction variables, such as the width in $\eta$ of the second
leading jet, provide additional separation of light quark jets and
heavy flavor jets. Angular variables such as the cosine between
lepton and leading jet, or the $\phi$ difference 
between lepton and $\met$, separate FCNC interactions from all backgrounds. Reconstruction of the 
top quark by combining the $W$~boson with one of the jets discriminates against the $W$+jets background.
The top quark mass reconstructed with the leading jet separates FCNC signal events (where the leading 
jet typically comes from the top quark decay) from all backgrounds including \ttbar~(where the
leading jet comes from one of the two top quark decays).
~

\begin{figure}[!h!tbp]
\centering
\includegraphics[width=0.238\textwidth]{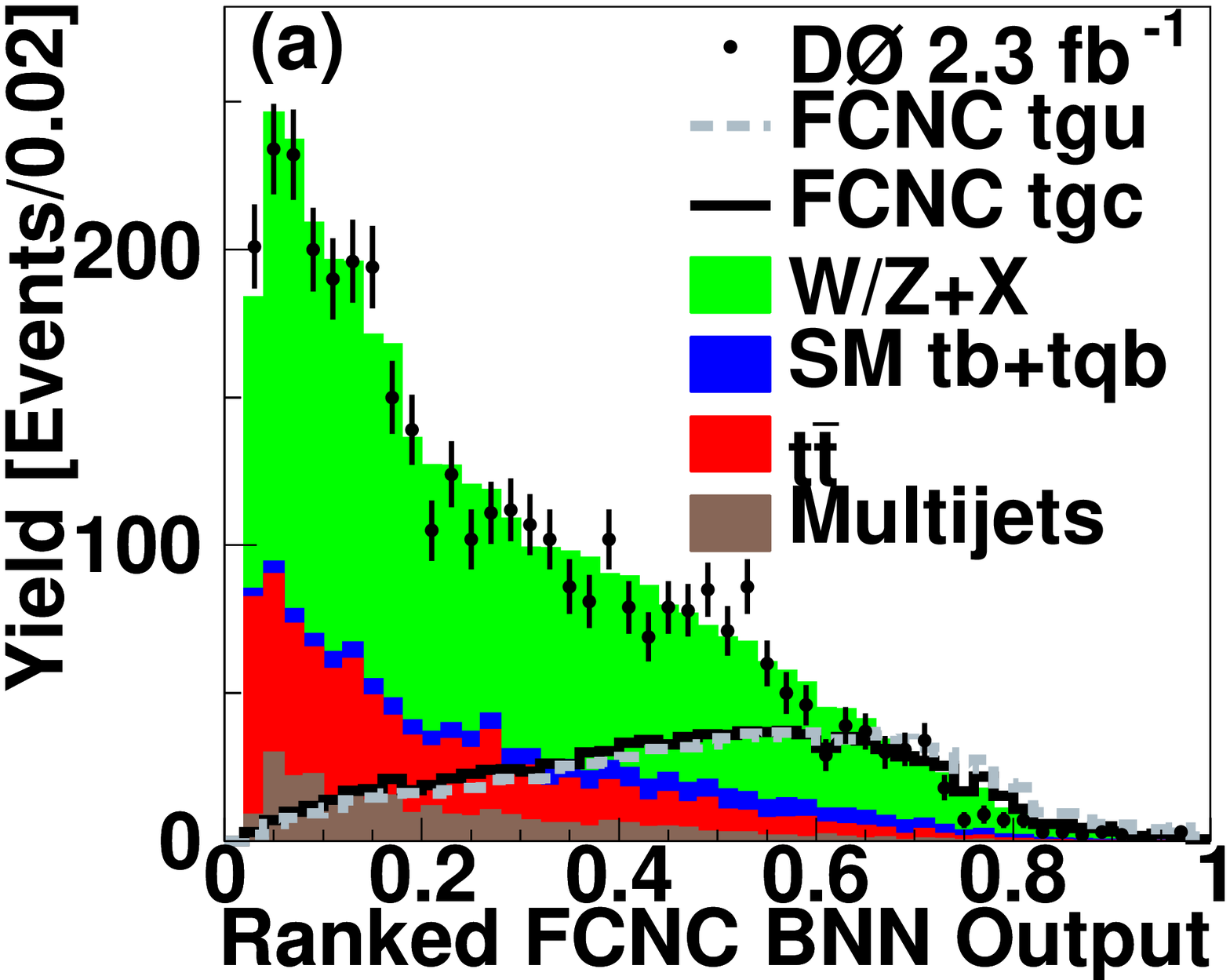}
\includegraphics[width=0.238\textwidth]{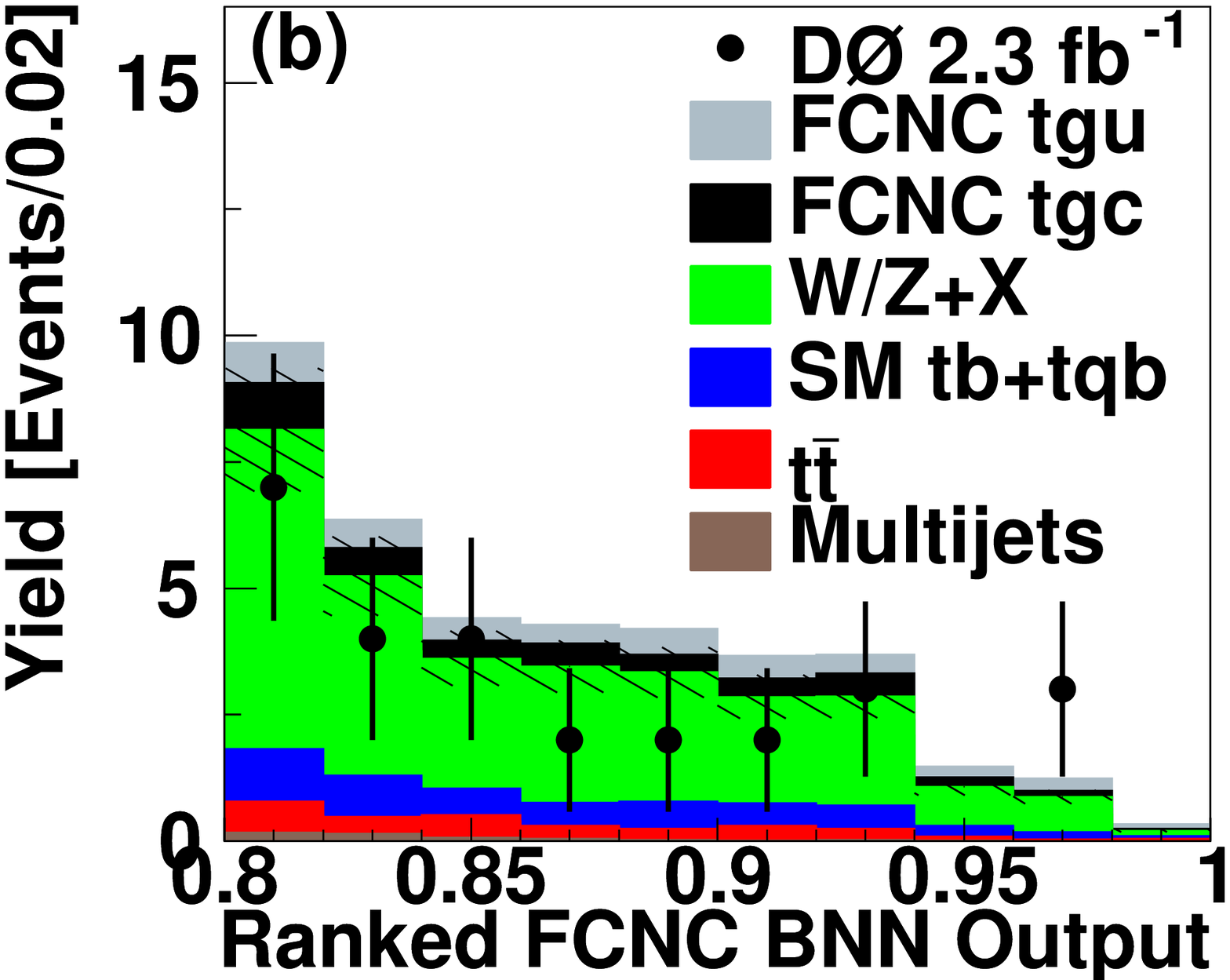}
\vspace{-0.1in}
\caption{Comparison of the background model to data for the FCNC discriminant summed over all
analysis channels, (a) for the whole discriminant range and (b) only the high discriminant region,
where the hatched region gives the uncertainty on the background sum. The bins have been ordered by their 
signal to background ratio and the FCNC signals are each normalized to a cross section of 5~pb in (a) and to 
their observed limits in (b).
$W/Z+X$ includes $W$+jets and smaller backgrounds from $Z$+jets and dibosons.
\label{fig:FCNCcombined} }
\vspace{-0.05in}
\end{figure}

Since their kinematics
are similar, the two FCNC processes are combined into a single signal for training the BNN, each normalized 
to the same coupling. Separate BNNs are trained for each choice of lepton flavor (electron or muon), jet 
multiplicity (2, 3, or 4), and data-taking period, twelve in total. Each utilizes 23 or 24 variables,
selected from the list of 54,  to provide the highest sensitivity for each analysis channel.
Figure~\ref{fig:FCNCcombined} shows the comparison
between background and data for all twelve BNN discriminants combined. A Kolmogorov-Smirnov test comparing
the observed data to the background sum in Figs.~\ref{fig:FCNCcombined}(a) and~\ref{fig:FCNCcombined}(b) 
gives values of 0.38 and 1.0, respectively,
demonstrating that the background model reproduces the data well.
~

Systematic uncertainties on the modeling of signal and background are 
described in Ref.~\cite{d0-prd-2008}, with main uncertainties being from corrections to
the jet energy scale and the $b$-tag modeling, with smaller contributions arising from
MC statistics, corrections for jet-flavor composition in $W$+jets events, 
and from the normalization of $W$+jets, multijets, and {\ttbar} production. The total uncertainty on the background is
(8--16)\%, depending on the analysis channel. For jet energy scale, $b$-tag modeling and $W$+jets
modeling, we vary not only the normalization but also consider effects on the shape of the final 
discriminant. When setting limits on the FCNC couplings, an additional signal cross section
uncertainty of 8.8\% from the NLO calculation is included~\cite{Liu}.

\begin{figure}[!h!tbp]
\includegraphics[width=0.35\textwidth] {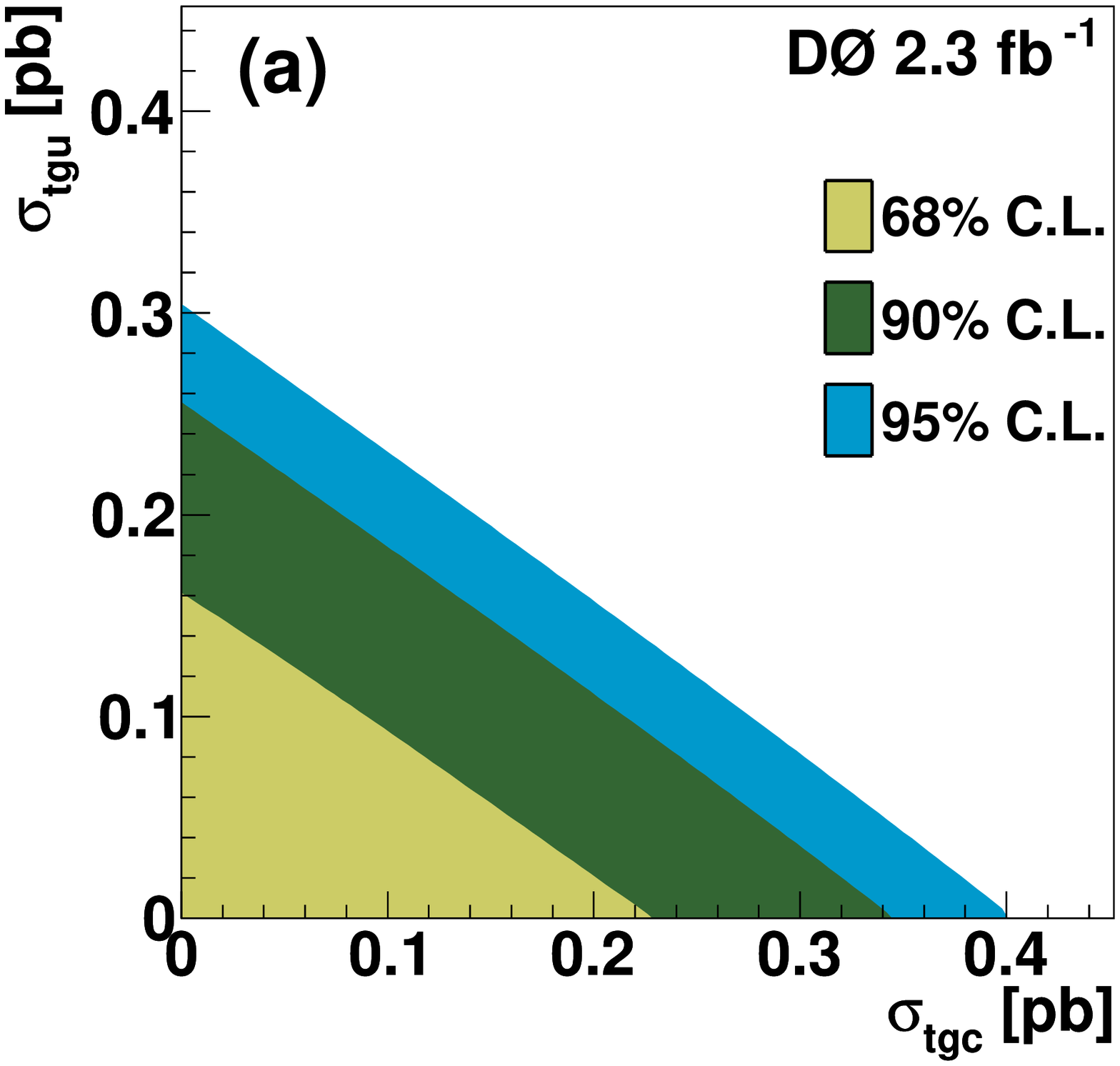}
\includegraphics[width=0.35\textwidth] {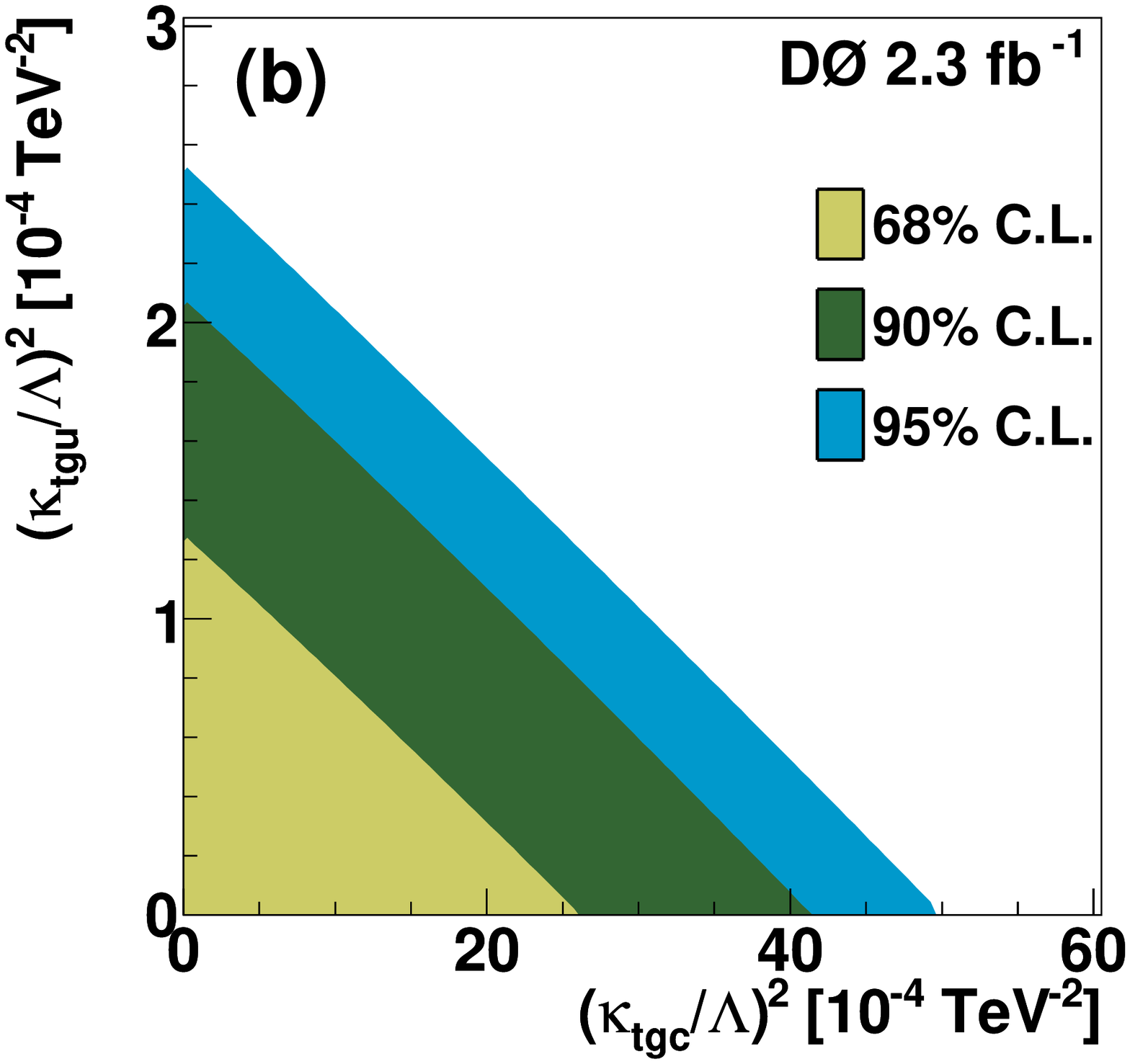}
\caption{Bayesian posterior probability as a function of (a) the $\sigma_{tgu}$ and $\sigma_{tgc}$
cross sections and (b) the squares of the couplings.
\label{fig:2dposterior} }
\end{figure}

Since the data are consistent with the background expectation, we 
set upper limits on the FCNC cross sections and couplings
using a Bayesian approach~\cite{bayes-limits}.
Following the analysis strategy of our previous work~\cite{Abazov:2007ev}, 
we form a two-dimensional Bayesian posterior density for the cross sections and for the
square of the FCNC couplings, using the BNN distributions for data,
background, and signals. Systematic 
uncertainties are taken into account with Gaussian priors, including correlations among bins and
signal and background sources. We choose priors that are flat and non-negative in the FCNC couplings squared 
and hence in the FCNC cross sections.
The posterior density as a function of the FCNC cross sections  
$\sigma_{tgu}$ and $\sigma_{tgc}$ is shown in Fig.~\ref{fig:2dposterior}(a).
We similarly form a two-dimensional Bayesian posterior density as a 
function of the $(\kappa_{tgf}/\Lambda)^2$, as shown in Fig.~\ref{fig:2dposterior}(b),
adding systematic uncertainties to the FCNC cross sections.
~

\begin{figure}[!h!tbp]
\includegraphics[width=0.35\textwidth] {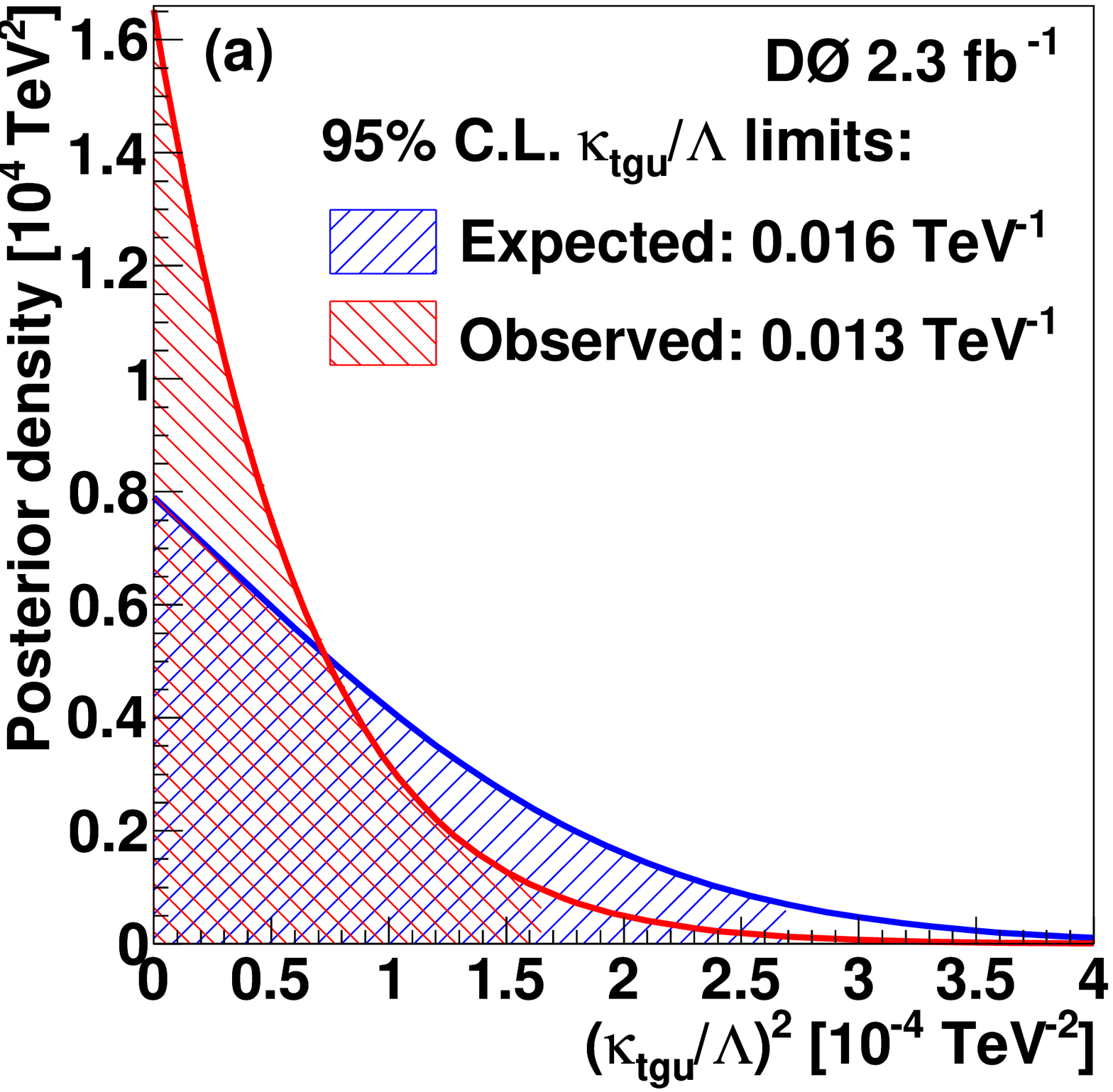}
\includegraphics[width=0.35\textwidth] {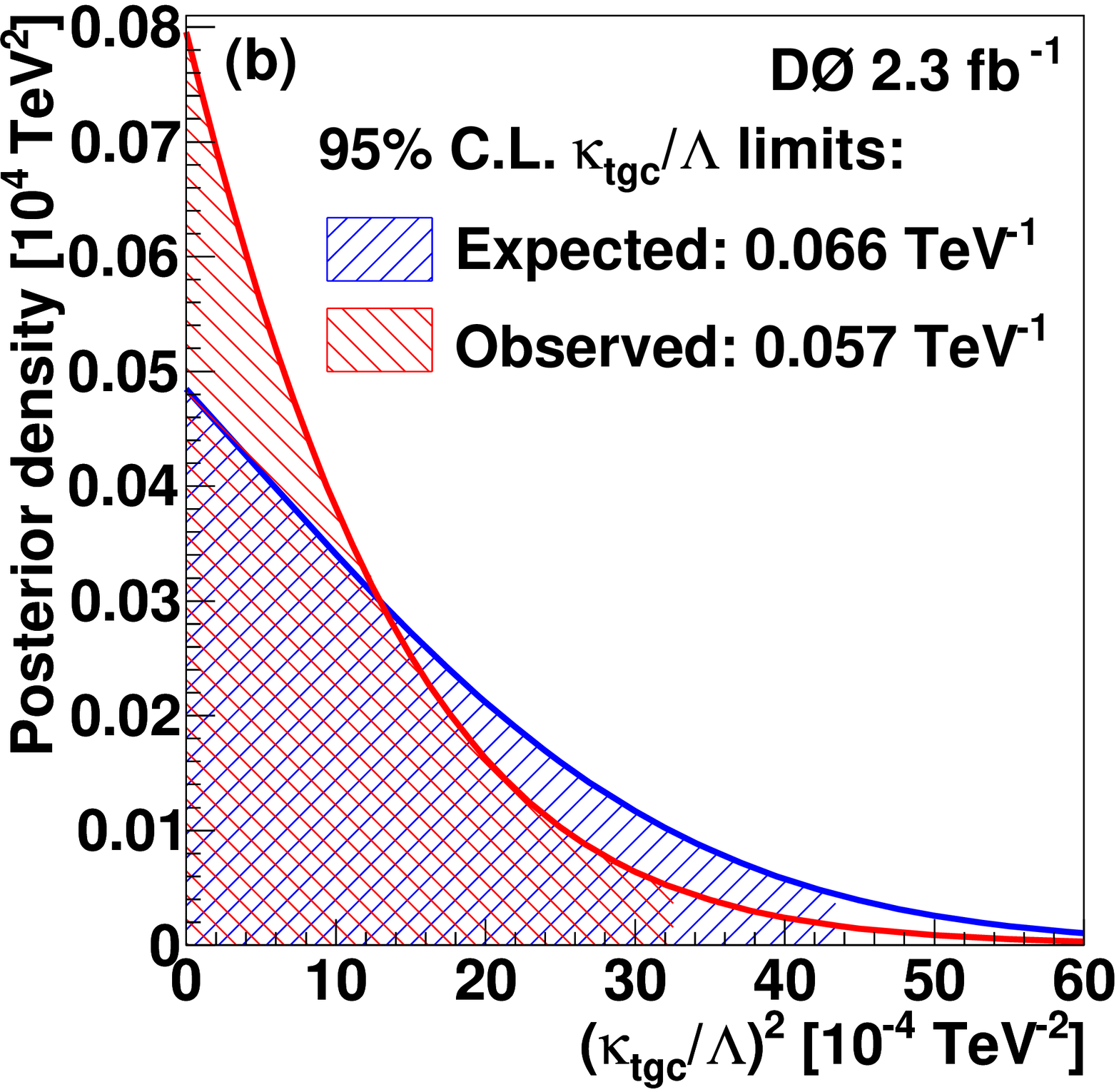}
\caption{One-dimensional Bayesian posterior probability as a function of 
(a) $(\kappa_{tgu}/\Lambda)^2$
and (b) $(\kappa_{tgc}/\Lambda)^2$.
\label{fig:1dposterior} }
\end{figure}

One-dimensional posterior densities as a function of 
$\sigma_{tgu}$ and $\sigma_{tgc}$ are derived from the
general two-dimensional posterior, by integrating over the $\sigma_{tgc}$ 
or $\sigma_{tgu}$ axes, respectively. One-dimensional posteriors are similarly derived
as a function of $(\kappa_{tgu}/\Lambda)^2$ and $(\kappa_{tgc}/\Lambda)^2$ and are shown in 
Fig.~\ref{fig:1dposterior}. This procedure keeps the measurement
free of theoretical assumptions concerning the relationship between the two FCNC cross sections and 
couplings. 
~
~
For each quantity, we also compute expected limits by replacing the count in data in each bin 
by the background sum. The expected posterior densities for $(\kappa_{tgu}/\Lambda)^2$ and 
$(\kappa_{tgc}/\Lambda)^2$ are also shown in Fig.~\ref{fig:1dposterior}, together with the 95\% C.L. limits.
The observed limits are below the expected limits, consistent with Fig.~\ref{fig:FCNCcombined}(b), 
which shows that the data count is below the background expectation for several bins in the high BNN 
output region. 
Since the FCNC decay branching fraction is proportional to the square of the coupling, 
the limits on the couplings can be translated into decay branching
fraction limits based on the NLO calculation~\cite{Zhang:2008yn}.
The limits on cross sections, couplings and branching fractions are summarized in 
Table~\ref{tab:limits}. 
~
\begin{table}[!h!btp]
\centering
\vspace{-0.15in}
\caption{Observed 95\% C.L. upper one-dimensional limits on the FCNC cross sections, couplings,
and branching fractions.}
\label{tab:limits}
\begin{tabular}{lcc}
\hline
\hline
                                 & $tgu$                & $tgc$ \\
\hline
Cross section                    & 0.20 ~pb           & 0.27~pb \\
$\kappa_{tgf}/\Lambda$               &  \hspace{10pt} 0.013~TeV$^{-1}$ \hspace{10pt}   &  \hspace{10pt} 0.057~TeV$^{-1}$ \hspace{10pt} \\
$\mathcal{B}(t \rightarrow fg)$ & $2.0\times10^{-4}$ & $3.9\times10^{-3}$ \\
\hline
\hline
\end{tabular}
\vspace{-0.1in}
\end{table}

In summary, we have presented a search for FCNC interactions in the gluon coupling of top quarks to 
up~quarks or charm quarks. Using a sample of 2.3~fb$^{-1}$ of integrated luminosity 
recorded by the D0 experiment at the Tevatron Collider at Fermilab,  
we set limits on the couplings of  $\kappa_{tgu}/\Lambda < 0.013$~TeV$^{-1}$ and
$\kappa_{tgc}/\Lambda < 0.057$~TeV$^{-1}$, without making assumptions about the $tgc$ and $tgu$ couplings,
respectively. The corresponding limits on top quark decay branching fractions
are $\mathcal{B}(t \rightarrow gu)<2.0\times10^{-4}$ and $\mathcal{B}(t \rightarrow gc)<3.9\times10^{-3}$.
These branching fraction limits are the most stringent and improve on the previous best limits by factors 
of two for $\mathcal{B}(t \rightarrow gu)$ and 1.5 for $\mathcal{B}(t \rightarrow gc)$~\cite{Aaltonen:2008qr}.
They improve on D0's previous result by a factor eight as a result of a larger data set
and significant improvements in analysis~\cite{Abazov:2007ev}.

%
We thank the staffs at Fermilab and collaborating institutions,
and acknowledge support from the
DOE and NSF (USA);
CEA and CNRS/IN2P3 (France);
FASI, Rosatom and RFBR (Russia);
CNPq, FAPERJ, FAPESP and FUNDUNESP (Brazil);
DAE and DST (India);
Colciencias (Colombia);
CONACyT (Mexico);
KRF and KOSEF (Korea);
CONICET and UBACyT (Argentina);
FOM (The Netherlands);
STFC and the Royal Society (United Kingdom);
MSMT and GACR (Czech Republic);
CRC Program and NSERC (Canada);
BMBF and DFG (Germany);
SFI (Ireland);
The Swedish Research Council (Sweden);
and
CAS and CNSF (China).

\vspace{-0.1in}



\begin{thebibliography}{99}
\bibitem{etaphi}
Pseudorapidity is defined as $\eta = -\ln[\tan(\theta/2)]$, where
$\theta$ is the polar angle with respect to the beam axis and 
the origin at the interaction vertex.
\end{thebibliography}
\end{document}